\documentclass[twocolumn,tighten]{aastex63}
\usepackage{amssymb, amsmath,framed}

\usepackage{bm}
\expandafter\ifx\csname package@font\endcsname\relax\else
 \expandafter\expandafter
 \expandafter\usepackage
 \expandafter\expandafter
 \expandafter{\csname package@font\endcsname}
\fi
\def\bq{\begin{equation}}
\def\eq{\end{equation}}
\def\bqy{\begin{eqnarray}}
\def\eqy{\end{eqnarray}}

\def\p{\partial}

\def\p{\partial}

\def\R{\mathbb{R}}

 \def\q#1#2{q^{\hspace{1pt}#1}_{\,,\hspace{1pt}#2}}
 \def\a#1#2{a^{\hspace{1pt}#1}_{\,,\hspace{1pt}#2}}
 \def\d#1#2{\delta^{\hspace{1pt}#1}_{\,,\hspace{1pt}#2}}

\begin{document}
\title{\large{On the Habitable Lifetime of Terrestrial Worlds with High Radionuclide Abundances}}

\correspondingauthor{Manasvi Lingam}
\email{mlingam@fit.edu}

\author{Manasvi Lingam}
\affiliation{Department of Aerospace, Physics and Space Sciences, Florida Institute of Technology, Melbourne FL 32901, USA}
\affiliation{Institute for Theory and Computation, Harvard University, Cambridge MA 02138, USA}

\author{Abraham Loeb}
\affiliation{Institute for Theory and Computation, Harvard University, Cambridge MA 02138, USA}

\begin{abstract}
The presence of a liquid solvent is widely regarded as an essential prerequisite for habitability. We investigate the conditions under which worlds outside the habitable zones of stars are capable of supporting liquid solvents on their surface over geologically significant timescales via combined radiogenic and primordial heat. Our analysis suggests that super-Earths with radionuclide abundances that are $\gtrsim 10^3$ times higher than Earth can host long-lived water oceans. In contrast, the requirements for long-lived ethane oceans, which have been explored in the context of alternative biochemistries, are less restrictive: relative radionuclide abundances of $\gtrsim 10^2$ could be sufficient. We find that this class of worlds might be detectable ($10\sigma$ detection over $\sim 10$ days integration time at $12.8$ $\mu$m) in principle by the James Webb Space Telescope at distances of $\sim 10$ pc if their ages are $\lesssim 1$ Gyr.\\
\end{abstract}

\section{Introduction} \label{SecIntro}
Due to the rapid growth of exoplanetary science, there has been renewed interest in determining the conditions that render a planet habitable \citep{Kast12,Ling19}. One of the most widely used criteria for habitability is the existence of a suitable liquid solvent, based on which the concept of the habitable zone (HZ) was introduced. The HZ is defined as the region around the host star where liquid water can theoretically exist on the planet's surface \citep{Dole,KWR93,Ram18,RAF19}. It is, however, crucial to recognize that habitability is a broader concept than the HZ, implying that worlds outside the HZ may be habitable. 

It is expected that worlds beyond the HZ outnumber those in the HZ by orders of magnitude \citep{LL19}. The most widely studied class of potentially habitable worlds beyond the HZ, motivated by the likes of Europa and Enceladus in our Solar system, is icy worlds with subsurface oceans \citep{NP16}. Along similar lines, it has been proposed that carbonaceous asteroids and Kuiper belt objects could also be abodes for life \citep{AM11}. Looking even farther afield, free-floating planets might also be capable of hosting life, especially in subsurface environments \citep{Ste99,Bad11,AS11,LL19}. In many of these cases, the heating necessary for sustaining liquid solvents is derived from radioactivity, which is sensitive to the age of the world under consideration.

In this Letter, we investigate the role of primordial and radiogenic heating in sustaining liquid solvents on the surface as a function of the size, age and radionuclide abundances of the objects. Unlike most previous studies, we will analyze the prospects for multiple solvents because there are no definitive grounds for supposing that water is the only solvent capable of hosting life \citep{SMI18}. We also study the prospects for their detectability via JWST. 

\section{Contributions to surface heat flux}
The current heat flow to Earth's surface is generated by radiogenic decay and primordial heat from Earth's formation \citep{Mel11}. We can likewise estimate the corresponding heat fluxes for other worlds. Before doing so, we note that the total radiogenic surface heat flux ($\mathcal{Q}_R$) is estimated via
\begin{equation}
  \mathcal{Q}_R = \sum_i \mathcal{Q}_{i0}\,\exp\left(-\frac{t}{\tau_i}\right),  
\end{equation}
where $t$ denotes the age of the world, whereas $\mathcal{Q}_{i0}$ and $\tau_i$ represent the initial radioactive heat flux (at $t=0$) and characteristic decay timescale of the $i^{\mathrm{th}}$ element, respectively. In our subsequent analysis, we will address short- and long-lived radionuclides separately and focus on the dominant isotopes.\footnote{A third category (``medium-life'' isotopes), comprising the likes of plutonium-244 ($\tau \sim 115$ Myr) and lead-205 ($\tau \sim 25$ Myr), may have to be included for certain stellar systems \citep{LOK18}. For Solar system analogs, however, the conventional division into short- and long-lived nuclides is sufficient.}

First, we consider the contribution from the decay of long-lived radionuclides such as uranium-238 and thorium-232. The radiogenic heating rate (in W) is usually modeled as being proportional to the mass $M$ of the world; this amounts to assuming that the heating rate per unit mass (units of W/kg) is constant. Using this scaling in conjunction with the mass-radius relationship $R \propto M^{0.27}$ \citep{ZSJ16}, the radiogenic heat flux associated with long-lived radionuclides ($\mathcal{Q}_L$) is
\begin{equation}\label{LLRadFlux}
    \mathcal{Q}_L \approx 0.166\,\mathrm{W\,m^{-2}}\,\Gamma_L \left(\frac{M}{M_\oplus}\right)^{0.46} \exp\left(-\frac{t}{\tau_L}\right),
\end{equation}
where $\tau_L$ represents the characteristic timescale for the decay of long-lived radionuclides and $\Gamma_L$ measures the abundance of radionuclides per unit mass relative to the Earth. In obtaining this formula, we used the fact that radionuclide concentration exhibits exponential decay along with the scaling $\mathcal{Q}_L \propto M/R^2$. Although the magnitude of $\tau_L$ depends on the specific composition of the world under consideration, we will adopt $\tau_L \approx 3.61$ Gyr for Earth-like worlds (see Section 6.24 of \citealt{TS02}). The normalization has been chosen such that we obtain a heat flux of $47\,\mathrm{mW\,m^{-2}}$ for modern Earth \citep{Kam11}. 

The next major contribution is the primordial heat released as the world cools after its formation. The surface heat flux arising from this activity ($\mathcal{Q}_P$) is estimated from Eq. (7.200) of \citet{TS02}:
\begin{equation}\label{PrimFlux}
    \mathcal{Q}_P = - \frac{R \bar{\rho} \bar{C}}{3} \frac{d T_m}{dt},
\end{equation}
where $\bar{C}$ is the mean specific heat, $\bar{\rho}$ denotes the mean mass density and $dT_m/dt$ signifies the average rate of mantle cooling. By using the scaling $R \propto M^{0.27}$, we find that $\bar{\rho} \propto M/R^3 \propto M^{0.19}$. Estimating the temporal dependence of $d T_m/dt$ is complicated because it depends on several mantle properties, but we will suppose that it can be heuristically modeled using Eq. (7.211) of \citet{TS02}:
\begin{equation}
    \frac{dT_m}{dt} = - \mathcal{C} T_m^2,
\end{equation}
where $\mathcal{C}$ is a composite variable that depends solely on the radionuclide composition and rheology of the mantle; however, it does not depend on the mantle thickness, its fiducial viscosity and current rate of heat generation. As our analysis deals with rocky worlds akin to Earth, we will hold $\mathcal{C} \sim \mathcal{C}_\oplus$, where $\mathcal{C}_\oplus \approx 1.2 \times 10^{-14}$ K$^{-1}$ yr$^{-1}$ is the value for Earth \citep{TS02}. After solving the above differential equation and substituting the expression into (\ref{PrimFlux}), we arrive at
\begin{equation}\label{PFluxFin}
    \mathcal{Q}_P \approx 0.04\,\mathrm{W\,m^{-2}}\, \left(\frac{M}{M_\oplus}\right)^{0.46} \left[1 + 0.1 \left(\frac{t - t_\oplus}{\tau_L}\right)\right]^{-2},
\end{equation}
where $t_\oplus \approx 4.54$ Gyr is the Earth's current age; the normalization factor preserves consistency with present-day Earth \citep{Kam11}. 

Although $\mathcal{Q}_L$ and $\mathcal{Q}_P$ represent the dominant long-term contributions to the surface heat flow, it is important to recognize that short-lived radionuclides can contribute to brief, but intense, initial heating \citep{HS06,AM11}. Although a number of short-lived radionuclides exist, the most notable among them are iron-60 and aluminium-26, but we shall concentrate on the latter due to its relatively higher abundance in planetesimals of the early Solar system \citep{LGG16}. The heat flux associated with this short-lived radionuclide ($\mathcal{Q}_S$) is obtained by following the derivation that led to (\ref{LLRadFlux}), thereby yielding
\begin{equation}\label{SLRadFlux}
    \mathcal{Q}_S \approx 89.5\,\mathrm{W\,m^{-2}}\,\Gamma_S \left(\frac{M}{M_\oplus}\right)^{0.46} \exp\left(-\frac{t}{\tau_S}\right),
\end{equation}
where $\tau_S \approx 1.02$ Myr, $\Gamma_S$ measures the abundance of aluminium-26 relative to early Solar system planetesimals, and the normalization constant follows upon utilizing the heat generation per unit mass \citep{AM11}; this derivation assumed that the world is primarily composed of rocky material.

The total heat flux ($\mathcal{Q}$) is given by the sum of the various contributions, i.e., we have
\begin{equation}\label{Qtot}
  \mathcal{Q} = \mathcal{Q}_L + \mathcal{Q}_P + \mathcal{Q}_S.  
\end{equation}
We are now in a position to estimate the effective temperature ($T_\mathrm{eff}$) using the Stefan–Boltzmann law:
\begin{equation}\label{SBLaw}
    \mathcal{Q} = \sigma T_\mathrm{eff}^4.
\end{equation}
At this stage, it is important to recognize that $T_\mathrm{eff}$ is not only dependent on the age of the world but also on its mass as well as $\Gamma_L$ and $\Gamma_S$. In order to simplify the problem, we will presume that the relative abundances of short- and long-lived radionuclides are equal to each other, i.e., we specify $\Gamma_L = \Gamma_S = \Gamma$. 

\section{Surface temperature and habitability duration}

\begin{figure}
\includegraphics[width=7.5cm]{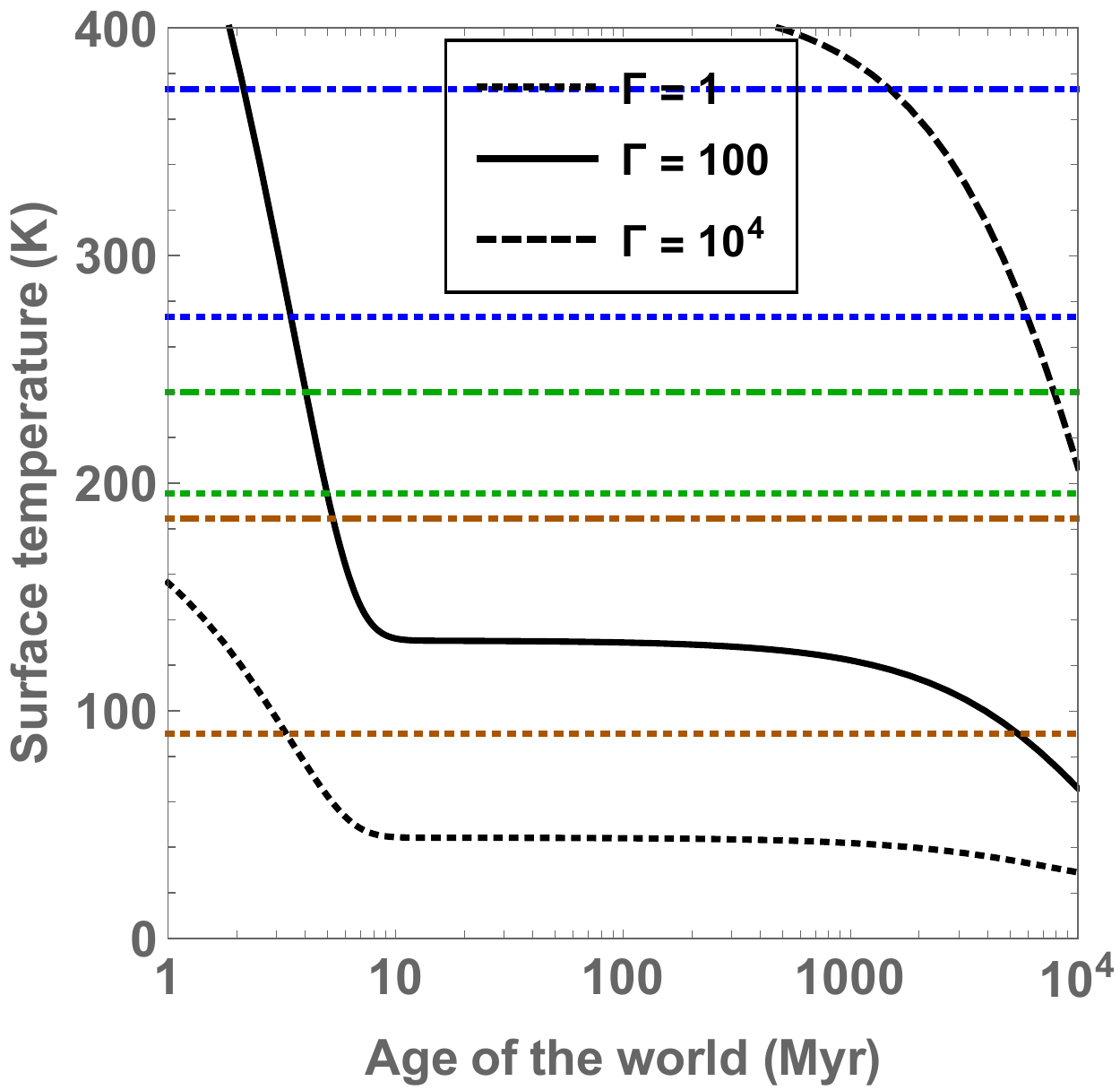} \\
\caption{Surface temperature (K) as a function of the age (in Myr) for an object with $M \approx M_\oplus$. The black curves correspond to different choices of radionuclide concentrations relative to Earth. The horizontal blue, green and brown lines delineate the regions where water, ammonia and ethane oceans can exist, respectively.}
\label{FigRadioHab1}
\end{figure}

Our goal is to determine the duration of \emph{surface} habitability, i.e., the length of time over which a liquid solvent can exist on the surface. There are two points that need to be underscored at this juncture. First, we do not address the prospects for subsurface habitability in this work. The primary reason is that subsurface biospheres are virtually impossible to detect at interstellar distances, even if they emit plumes \citep{LL19}. The temperature in the interior ($T_i$) can be calculated by solving the heat transfer equation (see Eq. (4.13) of \citealt{Mel11}):
\begin{equation}
    \frac{\partial T_i}{\partial t} = \nabla \cdot \left(\kappa \nabla T_i\right) + \frac{\mathcal{H}}{C_P},
\end{equation}
where $C_P$ is the heat capacity at constant pressure, $\mathcal{H}$ is the heating rate per unit mass, and $\kappa$ is the thermal diffusivity; the time-dependent boundary condition for this PDE is set by $T_i(r=R) = T_s$, where $T_s$ is the surface temperature. We will not solve this PDE because it pertains to \emph{subsurface} habitable conditions.

\begin{figure}
\includegraphics[width=7.5cm]{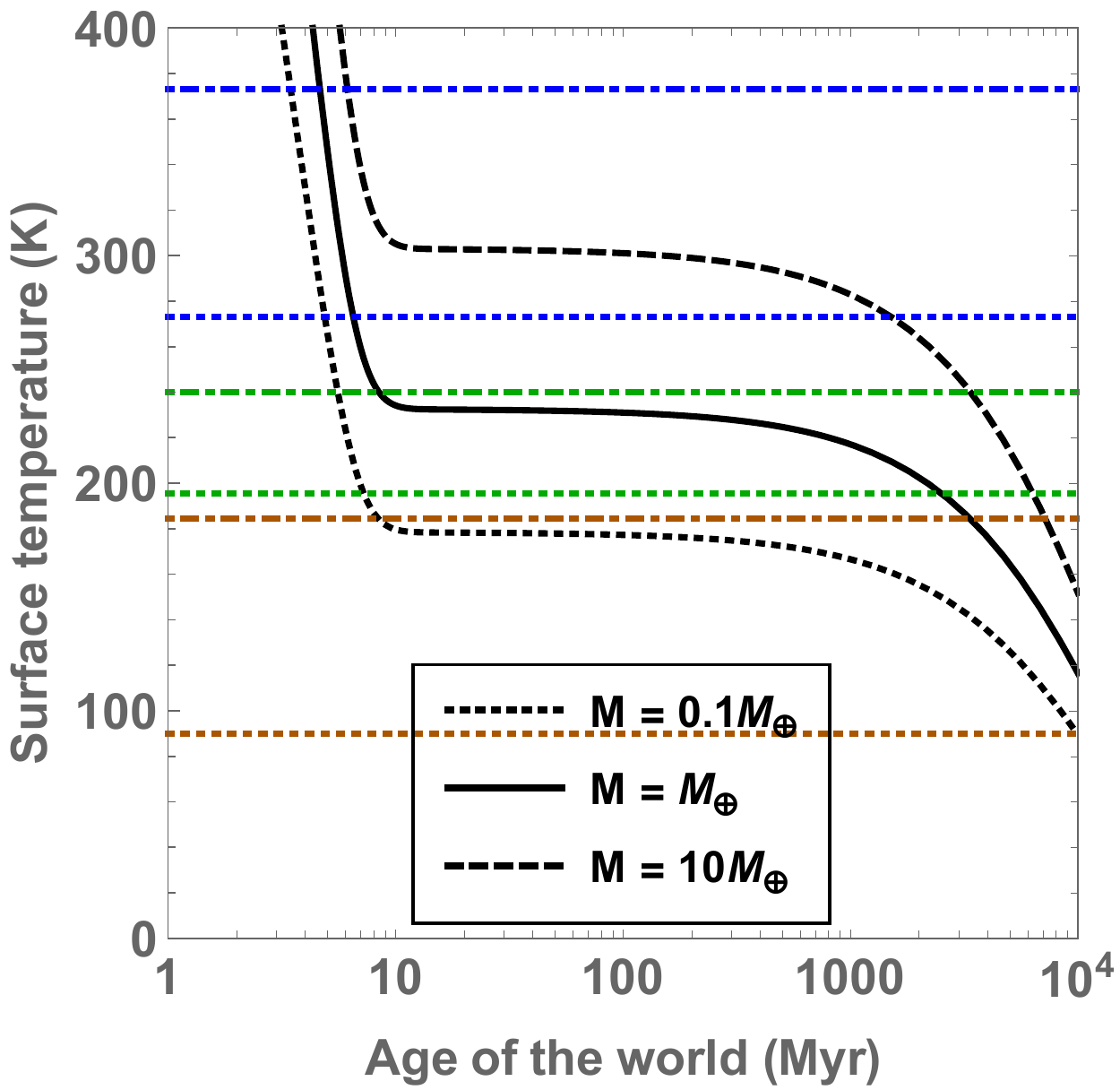} \\
\caption{Surface temperature (K) as a function of the age (in Myr) for a world with radionuclide abundances $\sim 10^3$ times higher than Earth. The black curves correspond to different choices of the object's mass. The horizontal blue, green and brown lines delineate the regions where water, ammonia and ethane oceans can exist, respectively.}
\label{FigRadioHab2}
\end{figure}

The second issue concerns the atmosphere, which permits the solvent to exist in liquid form on the surface. Massive and optically thick atmospheres have the capacity to mitigate cooling and thereby retain potentially habitable surface conditions for long timescales \citep{Ste99,Bad11}. We shall, instead, adopt the conservative choice of tenuous and optically thin atmospheres for several reasons. For starters, the habitability timescale herein constitutes a lower bound as explained above. In addition, biosignatures in thick atmospheres dominated by hydrogen and other reducing gases have not been widely modeled. Lastly, the retention of massive atmospheres over long timescales might prove to be difficult for some rocky worlds \citep{DLMC,Ling19}. For sufficiently tenuous atmospheres, the surface temperature ($T_s$) is approximated by $T_\mathrm{eff}$ hereafter. Our results for the habitability lifetime are not very sensitive to the optical depth if it is much smaller than unity.

To complete our analysis, it is necessary to choose appropriate solvents. A number of compounds apart from water have been identified as viable candidates for biochemical reactions to occur. We choose ammonia, an inorganic polar solvent, because it has been the subject of several experimental studies \citep{SMI18}. We adopt ethane, an organic polar solvent, on account of its presumed commonality and its potential to harbor alternative biochemistries \citep{BFM}. For a surface pressure of $\sim 1$ bar, water, ammonia and ethane are liquids between $\sim 273$ K to $\sim 373$ K, $\sim 195.5$ K to $\sim 240$ K and $\sim 90$ K to $\sim 184.5$ K, respectively. The surface temperature as a function of time is determined from (\ref{SBLaw}), and the above ranges are employed to estimate the habitability timescale ($t_H$).

Fig. \ref{FigRadioHab1} depicts the surface temperature for $M \approx M_\oplus$ and different choices of $\Gamma$. It is apparent that $\Gamma$ has a major impact on $t_H$. For $\Gamma = 1$, we find that neither water nor ammonia oceans are feasible, although ethane oceans can persist for $\sim 10$ Myr. In contrast, for $\Gamma = 100$, short-lived water and ammonia oceans may exist for $\sim 1$ Myr, but ethane oceans are capable of surviving for several Gyr. Lastly, when it comes to $\Gamma = 10^4$, we observe that both water and ammonia oceans are feasible over Gyr timescales.

Next, we examine the role of the object's mass. From (\ref{Qtot}) and (\ref{SBLaw}), the ensuing scaling is $T_s \propto M^{0.115}$, thus suggesting that both $T_s$ and $t_H$ should have a weak dependence on $M$. In actuality, $t_H$ exhibits a complex relationship with $M$ due to the co-dependence on $\Gamma$. This becomes evident upon inspecting Fig. \ref{FigRadioHab2}, where $\Gamma = 10^3$ while $M$ is varied. For $M \approx 0.1 M_\oplus$, we observe that water and ammonia oceans can exist for only $\sim 1$ Myr. In contrast, for $M \approx 10 M_\oplus$, these oceans could persist over Gyr timescales.

\section{Discussion and Conclusions}
Finally, we will briefly discuss some implications for biology and feasibility of detection.

\subsection{Biological implications}
We do not have a clear picture of when life originated on Earth, under the implicit assumption that it was not transported via panspermia. As per the latest phylogenetic and paleontological evidence, allied to theoretical arguments, there are tentative grounds for supposing that abiogenesis occurred  $\lesssim 100$ Myr after Earth's formation \citep{LM94,DPG17,BPC18}. It is important, however, to recognize that we have no knowledge of the timescale for abiogenesis on other worlds \citep{ST12}. In addition, the process of terrestrial planet formation from the protoplanetary disk may necessitate $\lesssim 10$-$100$ Myr \citep{MLO12,Arm18}. 

Based on the above arguments, it would seem desirable for $t_H$ to exceed $100$ Myr in order for life to originate. If the pace of evolution is similar to that on Earth, the manifestation of detectable biosignatures engendered by microbes might require $\sim 1$ Gyr. Thus, in order for $t_H \sim 0.1$-$1$ Gyr to be realized, Figs. \ref{FigRadioHab1} and \ref{FigRadioHab2} suggest that $\Gamma \gtrsim 10^3$ and $M \gtrsim M_\oplus$ are necessary. This immediately raises the question of whether such high concentrations of long-lived radionuclides are attainable, and by what pathways they can occur. 

\begin{figure}
\includegraphics[width=7.5cm]{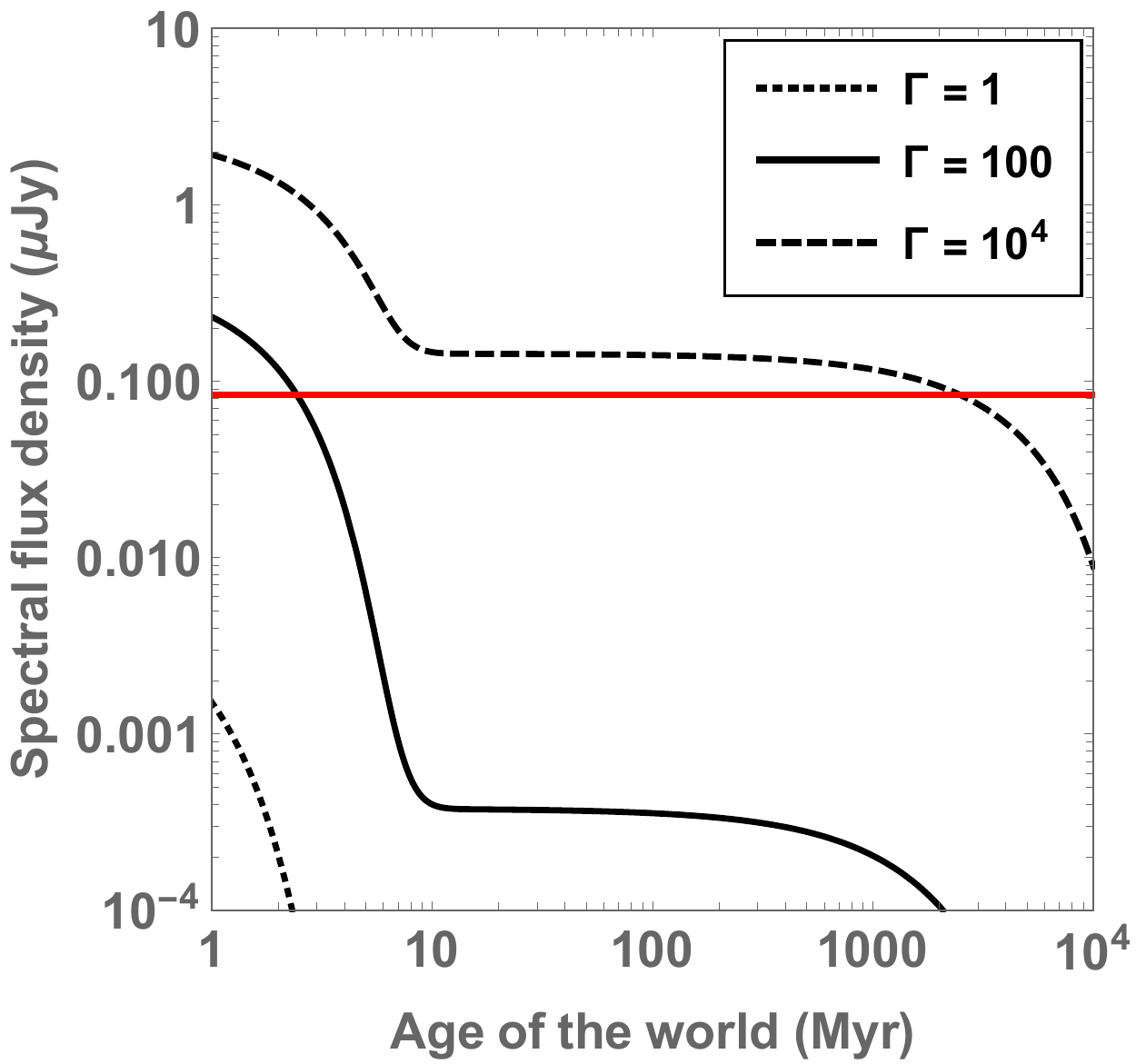} \\
\caption{Spectral flux density (in $\mu$Jy) as a function of the age (in Myr) for an object with $M \approx M_\oplus$ at a distance of $10$ pc. The black curves show different choices of radionuclide concentrations relative to Earth. The horizontal red line corresponds to the sensitivity of the MIRI-JWST for a $10\sigma$ detection over $\sim 10$ days at $12.8$ $\mu$m.}
\label{FigRadioHab3}
\end{figure}

Before tackling this issue, a comment on potential microbial ecosystems is in order. Although radioactivity levels will indubitably be high on these worlds, many Earth-based extremophiles (e.g., \emph{Deinococcus radiodurans}) readily tolerate high doses. In fact, the chemoautotrophic bacterium \emph{Desulforudis audaxviator} derives its energy requirements indirectly from radioactivity \citep{CBA08}. The surface heat flux is $\sim 300$-$1100$ W/m$^2$ when conditions suitable for liquid water exist. Of this energy flux, only a small fraction will be accessible to perform work due to thermodynamic constraints \citep{LiLo}. Nevertheless, even if merely $\sim 0.1\%$ of the total flux is utilized, it is still $\sim 4$ orders of magnitude higher than the basal requirement of $1.48 \times 10^{-5}$ W/m$^2$ imposed by metabolism \citep{KVW17}; it is also two orders of magnitude higher than the energy flux required by \emph{Chlorobiaceae} in the Black Sea \citep{MGK05}.

Now, we examine the abundances of long-lived radionuclides. Analysis of near-infrared signatures has revealed that r-process elements (which include uranium and thorium) are predominantly synthesized during neutron star mergers \citep{KMB17,KAB19}.\footnote{\url{https://blog.sdss.org/wp-content/uploads/2017/01/periodic_table.png}} Thus, if the rate of neutron star mergers per unit volume tracks the stellar density, it is conceivable that worlds in the inner regions of the Galactic bulge or in environments that are gas-poor (e.g., in elliptical galaxies) could exhibit higher radionuclide concentrations; furthermore, many metal-poor stars with enriched r-process elements have been detected \citep{Fre18}. 

Spectroscopic surveys of solar twins and analogs have shown that the solar abundance of thorium is a few times lower ($< 3$) than other stars in the sample \citep{UJP15,BMM19}. Overall, some worlds will be characterized by elevated abundances of long-lived radionuclides \citep{LOK18}. Turning our attention to short-lived radionuclides, specifically aluminium-26, they are unlikely to affect habitability over Gyr timescales, irrespective of their initial inventory. However, a number of crucial short-term factors depend on its abundance such as differentiation and dehydration of planetesimals \citep{GM09,LGB19}, and melting ice to initiate serpentinization \citep{GK17}. A number of metrics indicate that high levels of aluminium-26 are widespread in extrasolar systems, especially in young stars \citep{LOK18}. 

The situation is rendered quite different when we consider oceans of liquid ethane. For super-Earths with relatively modest enhancements of $\Gamma \gtrsim 10$, we find that $t_H \sim 0.1$-$1$ Gyr is realizable. Due to the relatively weaker constraints on $\Gamma$, the existence of long-lived ethane oceans is more plausible with respect to water oceans; for different reasons, \citet{BFM} concluded that ethane seas may be the most common in the Universe. Thus, there exist compelling grounds to investigate potential biochemistries in liquid ethane as well as the resultant biosignatures. 

If habitable conditions persist for $\gtrsim 100$ Myr, another factor must be taken into consideration for free-floating worlds. Due to the relative motion of stars as well as their ejection from the parent stellar system, they ought to have typical velocities of $\sim 20$ km/s with respect to nearby stars. Thus, over a period of $\sim 100$ Myr, they can traverse distances of $\gtrsim 1$ kpc. During the course of this journey, they may be captured by three-body gravitational interactions into binary systems \citep{GLL18}. Once they have been captured, this opens the possibility of seeding other worlds in that system via interplanetary panspermia.\footnote{Note that the prospects for interplanetary panspermia are enhanced for clustered planetary systems such as those around M-dwarfs \citep{LL17}.}

\subsection{Detectability of objects}

\begin{figure}
\includegraphics[width=7.5cm]{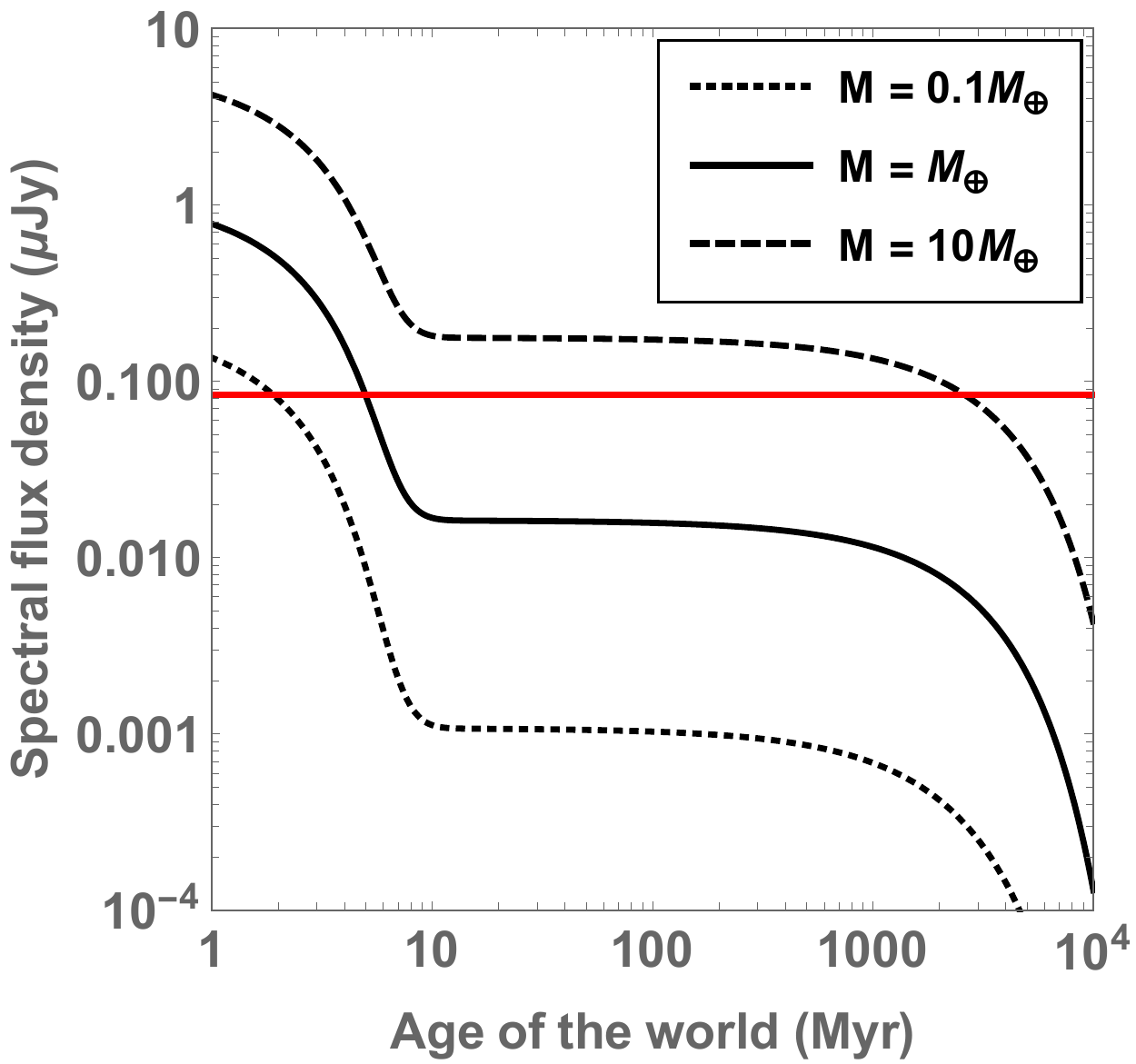} \\
\caption{Spectral flux density (in $\mu$Jy) as a function of the age (in Myr) for a world  at a distance of $10$ pc with radionuclide abundances $\sim 10^3$ times higher than Earth. The black curves embody different choices of the object's mass. The horizontal red line corresponds to the sensitivity of MIRI-JWST for a $10\sigma$ detection over $\sim 10$ days at $12.8$ $\mu$m.}
\label{FigRadioHab4}
\end{figure}

In order to detect worlds that are rendered habitable by radiogenic and primordial heating, the spectral flux density ($S$) received at Earth via black body emission must be sufficiently high (see \citealt{AS11}).\footnote{In some respects, our analysis is similar to the detection of free-floating Y-type brown dwarfs \citep{BSL03,Bil17}, because they have black body temperatures as low as $\sim 250$ K.} $S$ is determined by scaling the black body spectral radiance of the object as follows:
\begin{equation}
S = \frac{B_\lambda}{4} \left(\frac{R}{d}\right)^2 = \frac{h \nu^3 R^2}{2 c^2 d^2} \left[\exp\left(\frac{h \nu}{k_B T_\mathrm{eff}}\right) - 1\right]^{-1},
\end{equation}
where $B_\lambda$ is the Planck function, $\nu$ is the frequency and $d$ is the distance of the object from Earth. We adopt a wavelength of $12.8$ $\mu$m, for which the sensitivity of the imager corresponding to the Mid-Infrared Instrument (MIRI) instrument on board JWST is $0.84$ $\mu$Jy to achieve $10\sigma$ detection over $10^4$ s of integration time (see Table 2 of \citealt{RWB15}).\footnote{\url{https://www.jwst.nasa.gov/}} Hence, over an integration time of $\sim 10^6$ s (viz., $\sim 10$ days), the detection threshold would be lowered to $84$ nJy. At this wavelength, the above expression can be rewritten to yield
\begin{eqnarray}
 &&   S \approx 2.03\,\mathrm{\mu Jy}\left(\frac{M}{M_\oplus}\right)^{0.54}\left(\frac{d}{10\,\mathrm{pc}}\right)^{-2} \nonumber \\
&& \quad \quad \times\, \left[\exp\left(\frac{1124.82\,\mathrm{K}}{T_\mathrm{eff}}\right) - 1\right]^{-1}.
\end{eqnarray}
Due to the exponential dependence of $S$ on $T_\mathrm{eff}$, $S$ will naturally be very sensitive to the age, radionuclide abundance and mass of the object. 

Figs. \ref{FigRadioHab3} and \ref{FigRadioHab4} depict $S$ for the fiducial choices of $\lambda = 12.8$ $\mu$m and $d = 10$ pc. For relatively high radionuclide abundances ($\Gamma \gtrsim 10^3$) and high masses ($M \gtrsim M_\oplus$), we find that the MIRI instrument may detect such worlds provided they are $\lesssim 1$ Gyr old. In contrast, objects with lower values of $\Gamma$ and $M$ are unlikely to be detectable by MIRI unless they are located relatively close to Earth. Based on available estimates for the density of free-floating planets, \citet{LL19} suggested that the distance to the nearest roughly Mars-sized free-floating object might be on the order of $0.1$ pc. In this case, objects with $\Gamma \sim 100$ are potentially detectable by MIRI during the first $\sim 10$ Myr of their existence, while $\Gamma \gtrsim 10^3$ could boost detectability to Gyr timescales. 

Hence, in principle, planets with sufficiently high abundances that are relatively nearby might remain detectable over long timescales. A number of young stellar comoving groups have been detected at distances of $\sim 10$-$100$ pc from Earth with stellar ages of $\sim 10$-$50$ Myr \citep{ZS04}. Examples of such groups include the $\beta$ Pictoris moving group \citep{MB14} and the TW Hydrae association \citep{DTG14,GFMM}, whose median ages are $\sim 20$ Myr and $\sim 10$ Myr, respectively. If worlds far removed from the HZs of such stars are characterized by $M \gtrsim M_\oplus$ and sufficiently high radionuclide abundances, they may fulfill the requirements for detectability by JWST based on Figs. \ref{FigRadioHab3} and \ref{FigRadioHab4}. 

In the event that JWST detects either a free-floating object or one that is far beyond the HZ with an anomalously high temperature (and thin atmosphere), one can use $T_s$ to place constraints on its age and radionuclide abundance. For such worlds, thermal emission spectroscopy could be utilized to discern potential atmospheric biosignatures \citep{FAD18}.

\acknowledgments
We thank the reviewer for the constructive report. This work was supported in part by the Breakthrough Prize Foundation, Harvard University's Faculty of Arts and Sciences, and the Institute for Theory and Computation (ITC) at Harvard University.


\end{document}